\documentclass[apj]{emulateapj}
\usepackage{natbib}
\usepackage{amsmath}
\usepackage{amstext}
\usepackage{graphicx}

\usepackage{verbatim}

\newcommand{\dxt}{\Delta \bar{x}^2}
\newcommand{\dbtbt}{\delta B^2/B^2}

\newcommand{\dB}{\delta B}

\begin{document} 

\title{Energetic particle transport across the mean magnetic field:
  before diffusion} 

\shorttitle{Cross-field particle transport before
  diffusion}

\author{T. Laitinen}
\email{tlmlaitinen@uclan.ac.uk}
\author{S. Dalla}
\affil{Jeremiah Horrocks Institute, University of Central Lancashire,
  Preston, UK}

\shortauthors{Laitinen \& Dalla}

\begin{abstract}
  Current particle transport models describe the propagation of
  charged particles across the mean field direction in turbulent
  plasmas as diffusion. However, recent studies suggest that at short
  time-scales, such as soon after solar energetic particle (SEP)
  injection, particles remain on turbulently meandering field lines,
  which results in non-diffusive initial propagation across the mean
  magnetic field. In this work, we use a new technique to investigate
  how the particles are displaced from their original field lines, and
  quantify the parameters of the transition from field-aligned
  particle propagation along meandering field lines to particle
  diffusion across the mean magnetic field. We show that the initial
  decoupling of the particles from the field lines is slow, and
  particles remain within a Larmor radius from their initial
  meandering field lines for tens to hundreds of Larmor periods, for
  0.1-10~MeV protons in turbulence conditions typical of the solar
  wind at 1~AU. Subsequently, particles decouple from their initial
  field lines and after hundreds to thousands of Larmor periods reach
  time-asymptotic diffusive behaviour consistent with particle
  diffusion across the mean field caused by the meandering of the
  field lines. We show that the typical duration of the pre-diffusive
  phase, hours to tens of hours for 10~MeV protons in 1~AU solar wind
  turbulence conditions, is significant for SEP propagation to 1~AU
  and must be taken into account when modelling SEP propagation in the
  interplanetary space.
\end{abstract}


%\maketitle

\section{Introduction}\label{sec:introduction}

The propagation of cosmic rays through the heliosphere is affected by
the large-scale interplanetary magnetic field, and the turbulent
fluctuations superposed on it. Understanding the nature of the effect
of these fields on particle transport is necessary, as we want to
understand the sources and acceleration processes of different cosmic
ray populations.

The turbulent fluctuations in the interplanetary magnetic field can be
considered as scattering agents for the cosmic rays, prompting the
description of their propagation as random-walk. \citet{Parker1965}
used this concept to describe the propagation of cosmic rays in the
time-asymptotic limit as diffusion. Determining the connection between
the turbulence properties and the diffusion coefficients, however, has
proven to be a difficult task. \citet{Jokipii1966} considered a
quasi-linear approach, where the transport along the mean field
direction was affected by fluctuations of the scale of the particle's
Larmor radius, whereas the propagation across the mean field was
caused by the random walk experienced by the magnetic field lines due
to turbulent fluctuations. The field line random walk (FLRW) model has
since been extended to consider the compound effect of the particles
scattering along the random-walking field lines
\citep{Matthaeus2003,Shalchi2010a,RuffoloEa2012}, and the most
advanced models generally compare well with full-orbit simulations
\citep{GiaJok1999} and some cosmic ray observations
\citep{Burger2000}.

The particle cross-field diffusion has also been applied in modelling
solar energetic particle (SEP) propagation in the heliosphere
\citep[e.g.][]{Zhang2009,Droge2010, He2011, Giajok2012,
  Qin2013}. Recent SEP observations, however, have proved difficult to
reconcile with the models. The solar wind turbulence properties,
measured by spacecraft \citep[e.g.][]{Burlaga1976,Bavassano1982JGR}
and coupled with theoretical and modelling work, suggest that parallel
diffusion dominates over cross-field diffusion, with the diffusion
coefficient ratio $\kappa_\perp/\kappa_\parallel\sim 0.01$
\citep[e.g.][]{GiaJok1999,Burger2000,Potgieter2014}. However, fits of
SEP intensity profiles performed with injection scenario and diffusion
coefficients as free parameters suggest a considerably larger value,
$\kappa_\perp/\kappa_\parallel\sim 0.1$
\citep{Dresing2012,Droge2014}. On the other hand, the sharp dropouts
observed in some SEP events \citep[e.g.][]{Mazur2000} have been
considered as evidence of only negligible cross-field diffusion of
SEPs \citep{Droge2010,Wang2014}.

The problem of accounting for the observed fast cross-field
propagation for SEPs was recently addressed by
\citet{LaEa2013b}. Using full-orbit particle simulations, they found
that the initial cross-field propagation with respect to the mean
field direction is not diffusive, and can be described as
field-aligned propagation of particles along stochastically meandering
field lines. They concluded that for a uniform background magnetic
field with turbulence parameters corresponding to solar wind
conditions near Earth, 10~MeV protons propagated to distances of 1 AU
from the source remaining bound to their meandering field lines over
time-scales of 6 hours. However, relative to the mean magnetic field
direction, the meandering field lines spread the particles to a much
wider cross-field extent than the asymptotic diffusion assumption. At
later stage, the particles could be considered diffusive with respect
to mean field direction only after 20 hours from their injection.

The \citet{LaEa2013b} study thus indicated that for SEPs early in the
event, the use of the diffusion description for particle cross-field
propagation is invalid, and that only at longer time scales can its
use be justified. The question then arises: when and how does the
transition from non-diffusive to diffusive cross-field propagation
take place, and how is the transition related to properties of the
plasma turbulence?  How do the particles decouple from the field lines?

The particle decoupling from field lines has been discussed previously
in attempts to understand and develop a theory for the the
time-asymptotic cross-field diffusion of particles in turbulent
magnetic fields
\citep[e.g.][]{Qin2002apjl,Minnie2009,RuffoloEa2012}. However,
quantifying the process of the particles leaving their field lines
presents several challenges. The field line meandering is typically
much faster than the decoupling of a particle from a field line
\citep[e.g.][]{Fraschetti2011perptimetheory}. Thus, a particle's
displacement in the cross-field direction is a measure of the random
walk of the magnetic field line, rather than the particle's random walk
relative to the meandering field line. On the other hand,
determination of the particle's position relative to its original
field line suffers from the uncertainty due to the variation of the
magnetic field within the particle's path of gyration.

In this work, we introduce a new technique to determine the
cross-field displacement of a particle from the meandering magnetic
field line it initially follows. We use the new technique, presented in
Section~\ref{sec:models}, to quantify the process of particle
decoupling from its initial field line and evaluate contribution of
the decoupling of the particle to the propagation of the particles
across the mean magnetic field. In Section~\ref{sec:results} we show
that the particle propagation across the field can be divided in two
separate diffusion ranges, which are separated by a transition
range. In Section~\ref{sec:discussion}, we discuss the physical nature
of the diffusion phases and the transition phase between them, and
compare our results with current particle transport theories. Finally,
we draw our conclusions in Section~\ref{sec:conclusions}.

\section{Models}\label{sec:models}

\begin{figure}
  \centering
  \includegraphics[width=\columnwidth]{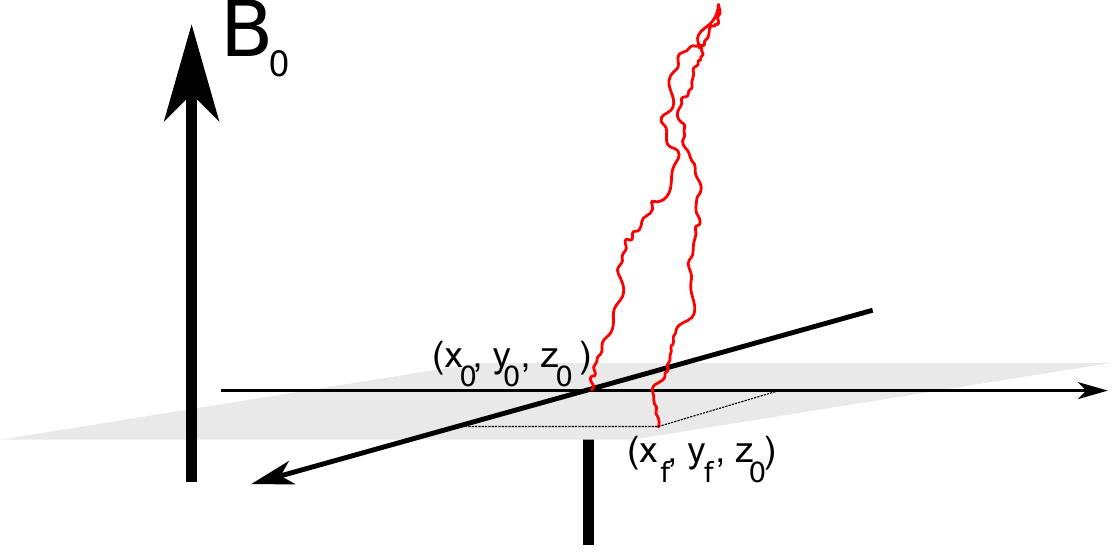}
  \caption{Schematic description of determining the cross-field
    displacement at $z=z_0$, with the particle's path shown by red
    curve. The particle starts at $(x_0,y_0,z_0)$ and is followed until
    it reaches the $z=z_0$ plane again, at $(x_f,y_f,z_0)$.\label{fig:zerocross}}
\end{figure}

We simulate charged particles by integrating their full orbits in a
magnetic field given by
\begin{equation}\label{eq:turbfield}
\mathbf{B}(x,y,z)=B_0 \hat{\mathbf{z}}+\delta \mathbf{B}(x,y,z),
\end{equation}
where $B_0$ is a constant background field, along the $z$-axis, and $
\delta \mathbf{B}(x,y,z)$ a fluctuating field, consisting of slab and
2D components, with energy ratio 20\%:80\% between the components, and a
broken Kolmogorov power law spectrum, with 
\begin{align}
  P_{\mathrm{slab}}(k_\parallel)&=\frac{\dB_\parallel^2}{B^2}\frac{C_\parallel}{1+(k_\parallel L_c)^{5/3}}\\
  P_{\mathrm{2D}}(k_\perp)&=\frac{\dB_\perp^2}{B^2}\frac{C_\perp}{1+(k_\perp L_c)^{8/3}}
\end{align}
where $\dB_\parallel^2$ and $\dB_\perp^2$ are the variances of the
turbulence slab and 2D components, respectively, $L_c$ is the
breakpoint scale of the turbulence, for which we use $L_C=2.15\,
\mathrm{R}_\odot$ in our study, with $\mathrm{R}_\odot$ the solar
radius, and $C_\parallel$ and $C_\perp$ are normalisation constants
\citep[see, e.g.,][]{GiaJok1999}. We use $B_0=5$~nT, consistent with
the magnetic field strength at 1 AU. The fluctuating field is formed
numerically as a sum of Fourier modes logarithmically spaced
  between wavenumbers $2\pi/(1 \mathrm{AU})$ and $2\pi/(10^{-4}
  \mathrm{AU})$, with the method described by
\citet{GiaJok1999}. The turbulence amplitude is parametrized by the
variance of the turbulence, $\dB^2$, which is varied in this study,
and the ratio between $\dB_\parallel^2$ and $\dB_\perp^2$, which is
20\%:80\% \citep{Gray1996} unless otherwise stated.

In this work, we are studying how the particles decouple from the
turbulent magnetic field lines. To measure this, we introduce a new
technique: we analyse the cross-field displacement of a particle that
returns back to the plane normal to the mean magnetic field that it was
injected at. The method is depicted in Figure~\ref{fig:zerocross}: The
particle is started at $(x_0,y_0,z_0)$ and traced until it returns
to the $z=z_0$ plane, where its coordinates $(x_f,y_f,z_0)$ are
recorded. A particle remaining perfectly on its field line would
return within two Larmor radii of its starting point. To eliminate
the displacement due to Larmor gyration, we calculate the
particle's gyrocenter $\bar{\mathbf{r}}$ at the injection and
return times, with
\begin{equation}
  \label{eq:gc}
  \bar{\mathbf{r}}=\mathbf{r}+\frac{q}{|q|\Omega B} \mathbf{v}\times\mathbf{B}
\end{equation}
where $\mathbf{r}$ and $\mathbf{v}$ are the particle's position and
velocity, respectively, and $q$ and $\Omega$ the particle charge and
gyrofrequency. As our model of turbulence is axisymmetric, either $x$
or $y$ can be used as representative direction perpendicular to the
mean field. We calculate the displacement in the $x$~direction, defined as
\begin{equation}
  \label{eq:gcdev2}
  \Delta \bar{x}(t)^2= (\bar{x}_f-\bar{x}_0)^2,
\end{equation}
where $t=t_f-t_0$ is the flight time of the particle, from the time of
particle injection at $t_0$ to its return to the $z_0$ plane at $t_f$,
and $\bar{x}_0$ and $\bar{x}_f$ are the $x$-coordinates of the
particle's gyrocenter at the start of they simulation and when it
returns to the $z=z_0$ plane, respectively. Defined in this way,
$\dxt$ does not include the cross-field propagation of the particles
directly due to the wandering of the field lines: were a particle to
follow the meandering field line precisely, its gyrocentre would cross
the starting plane at exactly the same location it started at,
$(\bar{x}_0,\bar{y}_0,\bar{z}_0)$, resulting in $\dxt=0$.

\section{Results}\label{sec:results}

\begin{figure}
  \centering
  \includegraphics[width=\columnwidth]{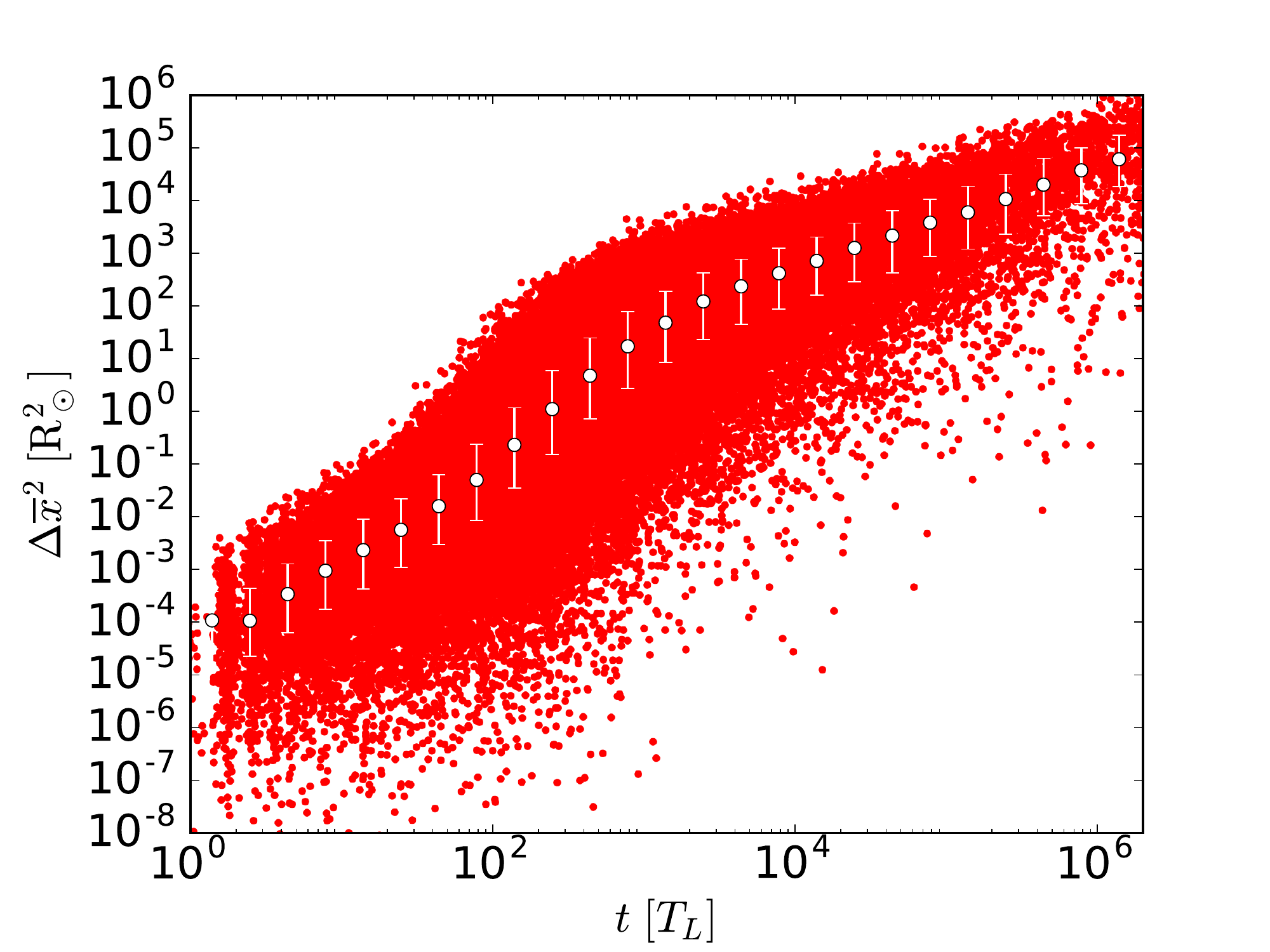}
  \caption{The displacement $\dxt$ of the returning
    particles as a function of time, in units $T_L=2\pi/\Omega$, for 10~MeV protons, with $\dbtbt=0.316$. The red symbols depict the
    displacement of each simulated particle, and the white symbols the
    median displacement for different times. The error bars are drawn at
    lower and upper quartiles.\label{fig:scatterplot_time}}
\end{figure}

\begin{figure*}
  \centering
  \includegraphics[width=.98\columnwidth]{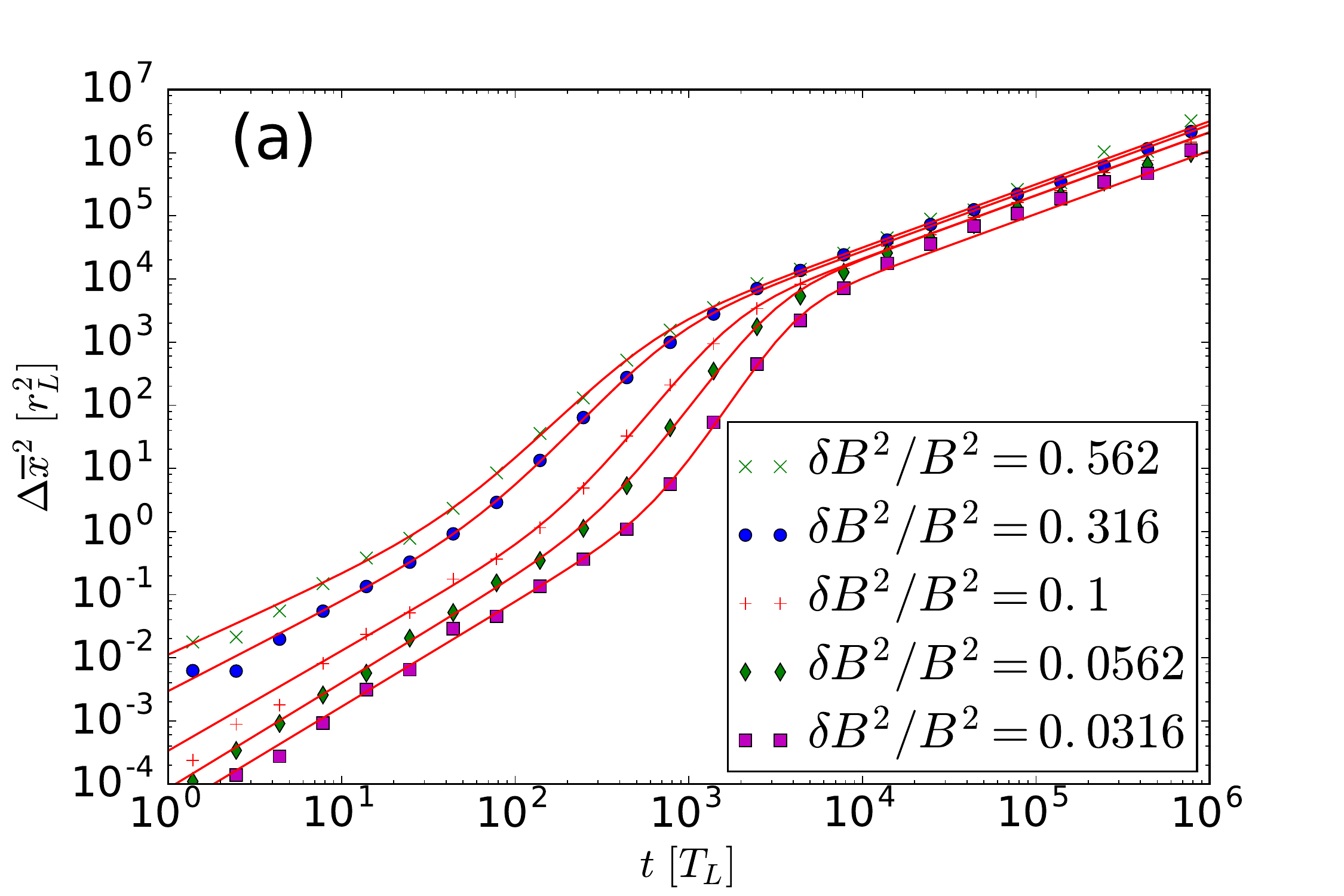}
  \includegraphics[width=.98\columnwidth]{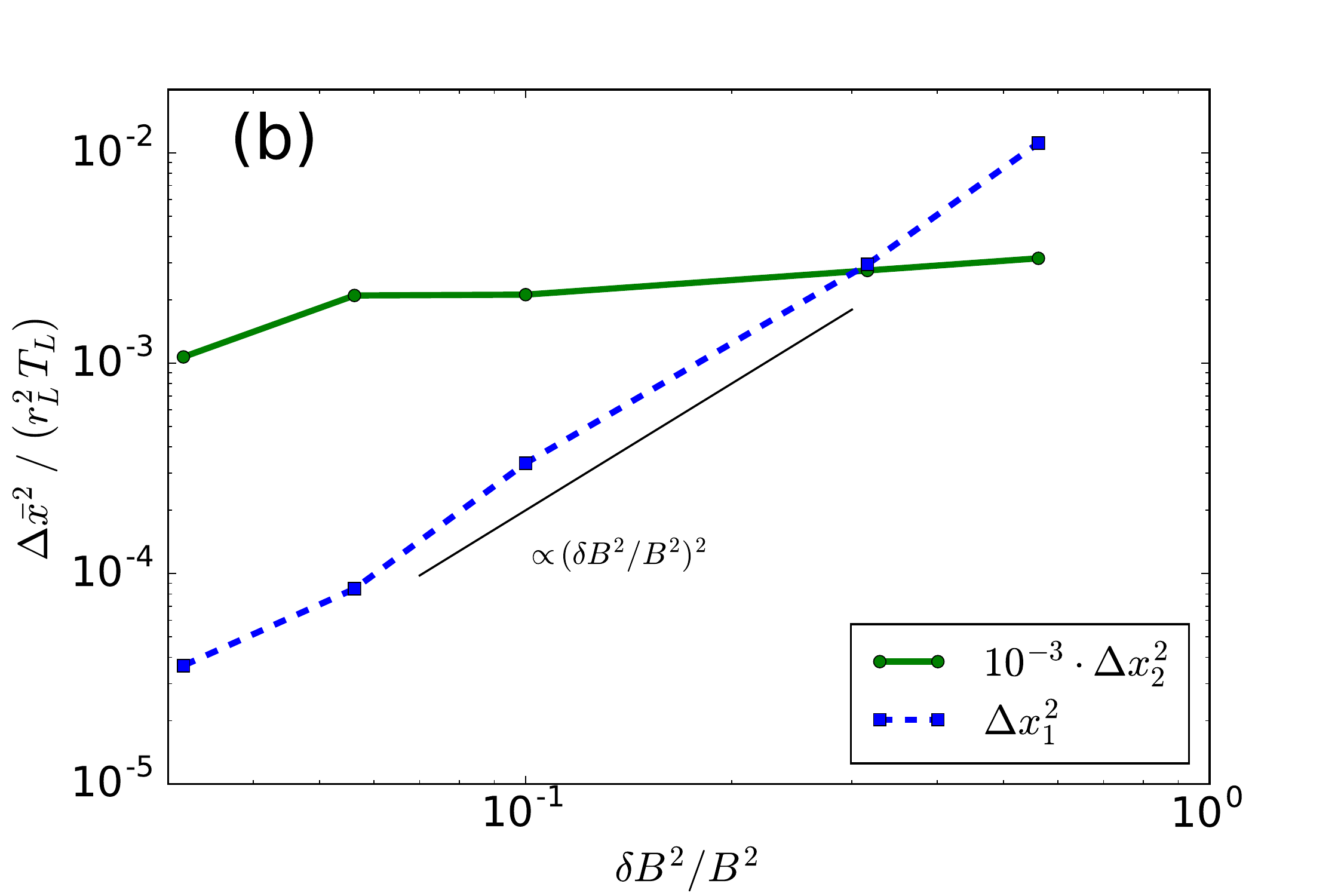}

  \includegraphics[width=.98\columnwidth]{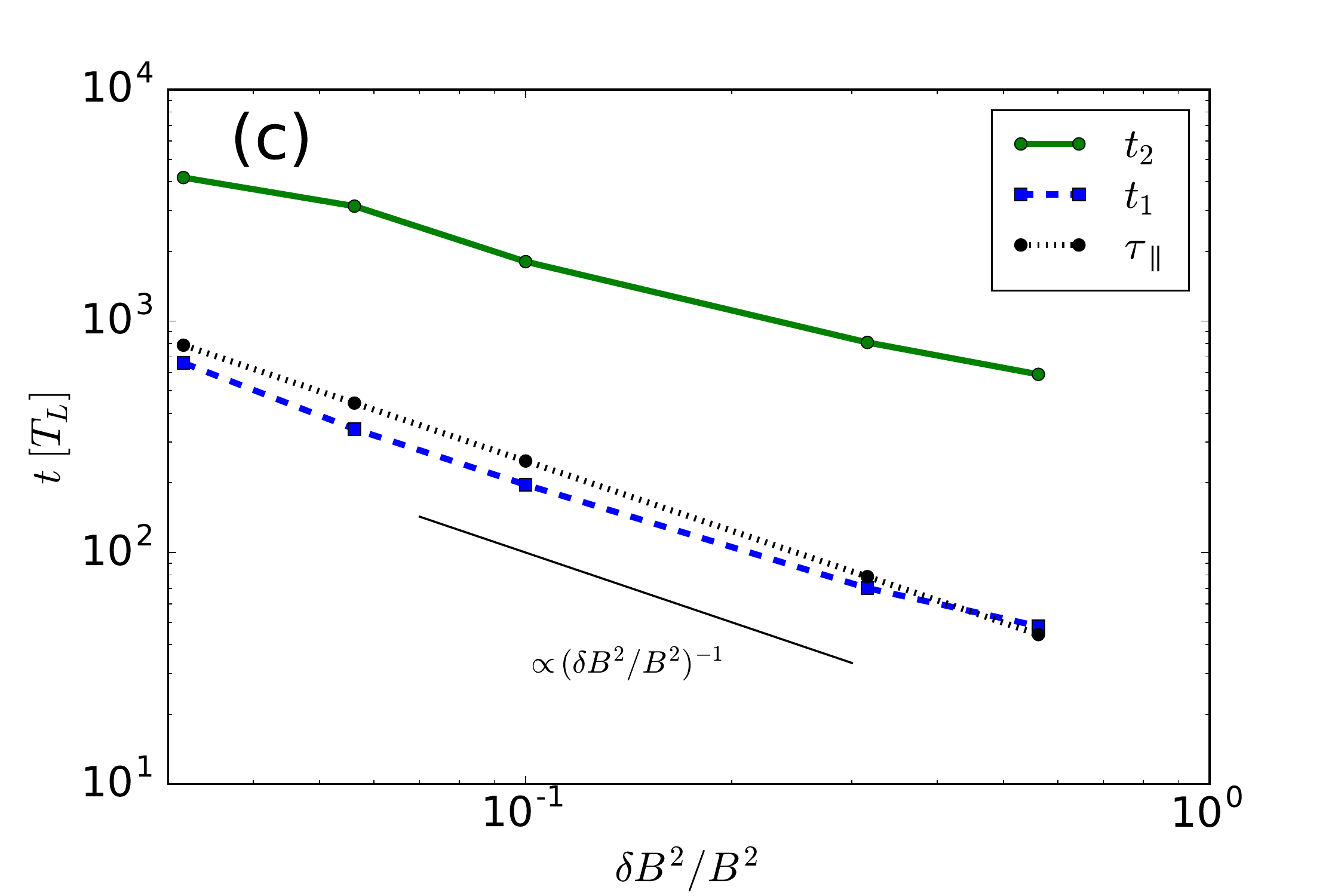}
  \includegraphics[width=.98\columnwidth]{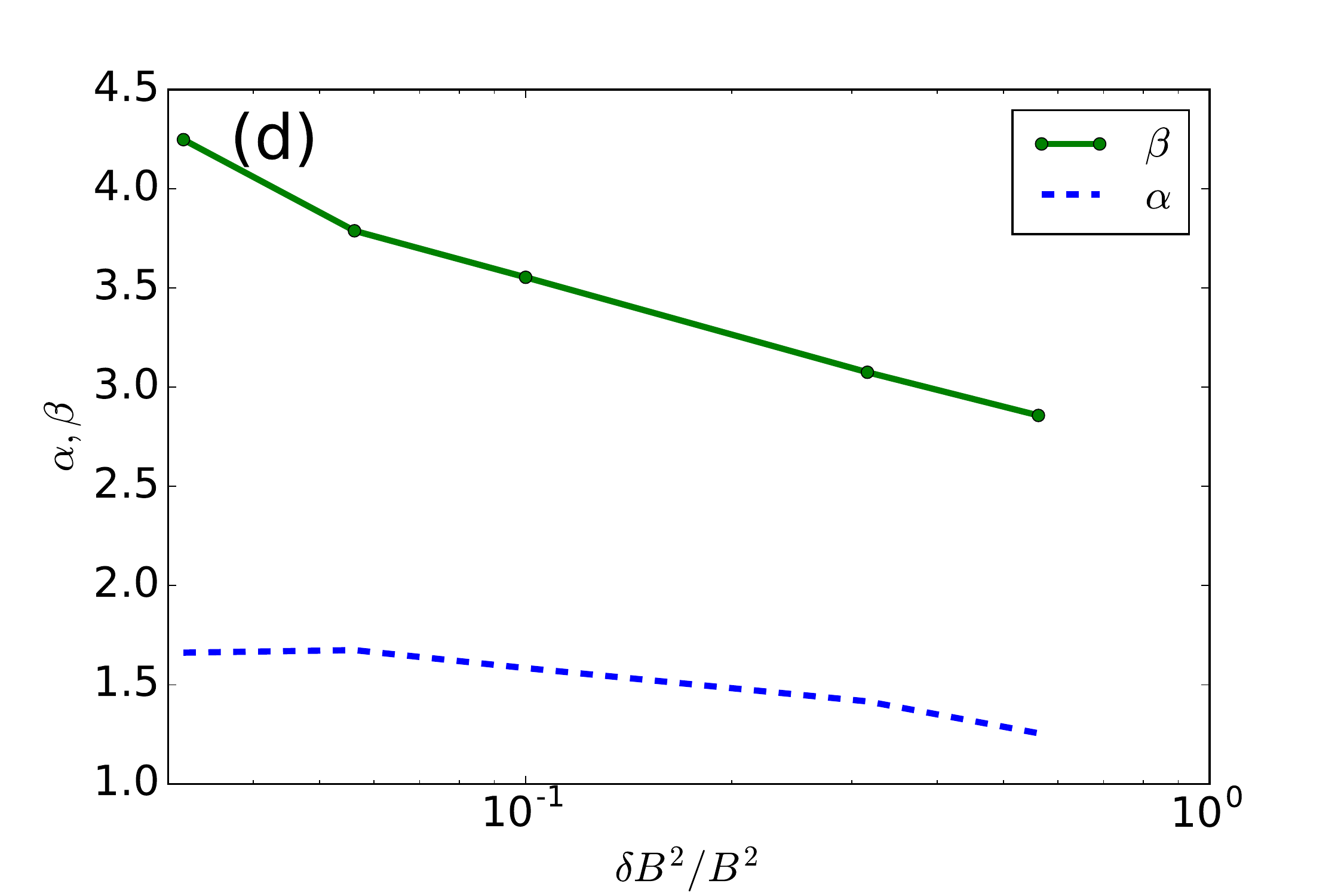}

  \caption{(a) Median displacement, in units $r_L^2$, versus time for
    five different $\dbtbt$, and the corresponding fit fo
    Equation~(\ref{eq:powerfit}). (b) $\dxt_1$ (dashed blue curve) and
    $10^{-3}\cdot\dxt_2$ (solid green curve) as a function of
    $\dbtbt$. (c) The transition times $t_1$ (dashed blue curve) and
    $t_2$ (solid green curve), and the parallel scattering timescale
    (black dotted curve), as function of $\dbtbt$. (d) The power
    law indices $\alpha$ (dashed blue curve) and $\beta$ (solid green
    curve) as function of $\dbtbt$. The thin black lines in panels (b) and (c) depict the trend lines discussed in the text.\label{fig:t_db2params}}
\end{figure*}

To analyse the cross-field propagation of energetic particles, we
studied the distribution of $\dxt(t_f)$, defined in
Equation~(\ref{eq:gcdev2}), within a monoenergetic population of
particles. The particles are injected at random locations
$(x_{0i},\,y_{0i},\,z_{0i})$ to minimise the possible effects of local
structures in the generated turbulent magnetic fields. We ran
simulations of typically 100,000 protons with isotropic pitch angle
distribution in the $v_z>0$ hemisphere. The particles were propagated in a turbulent magnetic
field until they returned back to initial plane, $z=z_{0i}$. At the
time of return, the square of the guiding centre displacement, as
given by Eq.~(\ref{eq:gcdev2}), was recorded.

We show an example of the simulation results in
Fig.~\ref{fig:scatterplot_time}, with a scatter-plot of $\dxt$ as a
function of the flight time $t$, for 10~MeV protons, with
$\dbtbt=0.316$. The time is normalised to the particle Larmor period,
$T_L=2\pi/\Omega$. The median displacement for logarithmically spaced
time ranges is shown by the white-filled circles, and the lower and
upper quartiles by the error bars. The Larmor radius of a
10~MeV proton in the given magnetic field is $0.13\,\mathrm{R}_\odot$,
thus for the particles for which $\dxt<0.017\,\mathrm{R}_\odot^2$, the
guiding centre of the returning particle remains within a gyroradius
of the initial location of the guiding centre.

We can identify three time ranges of different behaviour of $\dxt$ as
a function of flight time. The first range, up to $t \sim 100\, T_L$,
contains particles that return to the initial plane close to the
original location, and have roughly a linear trend of $\dxt$ as a
function of time, consistent with diffusive, or slightly
superdiffusive, increase of the displacement. At around $t\sim 100\,
T_L$, the spreading becomes faster, clearly super-diffusive, within
the second range. The fast spreading continues until at $t\sim 1000\,
T_L$ it relaxes back to a diffusive trend. For the purposes of this
study, we name these ranges {\em the first diffusion range}, {\em the
  transition range} and {\em the second diffusion range}.

In order to characterise the transition between the first and second
diffusion ranges, we must determine when the transition takes
place. To do this, we fit $\dxt$ as a function of time with a function
that depicts initially a non-diffusive behaviour, $\dxt\propto
t^\alpha$, followed by a fast spreading across the field with
$\dxt\propto t^\beta$, and a time-asymptotic diffusion, $\dxt\propto
t$. Overall, the function has the form
\begin{equation}\label{eq:powerfit}
  \dxt(t)=\dxt_1\;\left(\frac{t}{T_L}\right)^{\alpha}\;\frac{1+\left(t/t_1\right)^{\beta-\alpha}}{1+\left(t/t_2\right)^{\beta-1}},
\end{equation}
where $\dxt_1$ represents the square of displacement at $t=T_L\ll t_1 < t_2$,
and $t_1$ and $t_2$ are the start and end times of the transition
range, respectively. At early times, $t\ll t_1$, the equation
describes the first diffusion range, with
\begin{equation}\label{eq:firstdiff}
    \dxt(t)\xrightarrow{t\ll t_1} \dxt_1\;\left(\frac{t}{T_L}\right)^\alpha
\end{equation}
whereas at late times, $t\gg t_2$, the second diffusion range is given
as
\begin{equation}\label{eq:seconddiff}
    \dxt(t)\xrightarrow{t\gg t_2}\dxt_2\;\frac{t}{T_L}
\end{equation}
where
\begin{equation}\label{eq:dx1}
  \dxt_2= \dxt_1\;\frac{t_2^{\beta-1}}{t_0^{\alpha-1}\;t_1^{\beta-\alpha}}.
\end{equation}
It should be noted that the form of the fit function has no physical
justification as such. It is only used to trace the behaviour of the
particles in the three ranges.

\begin{figure*}
  \centering
  \includegraphics[width=.98\columnwidth]{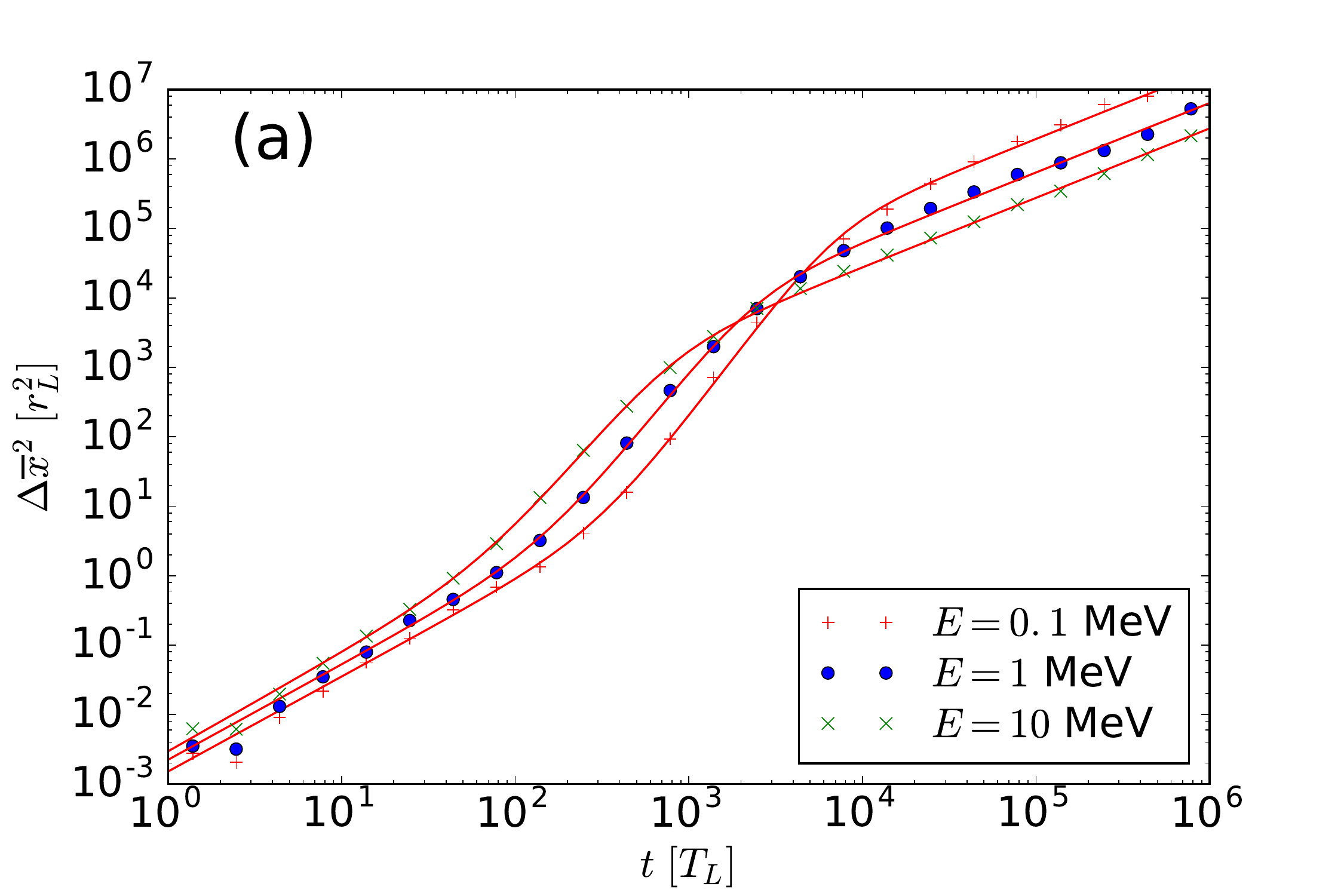}
  \includegraphics[width=.98\columnwidth]{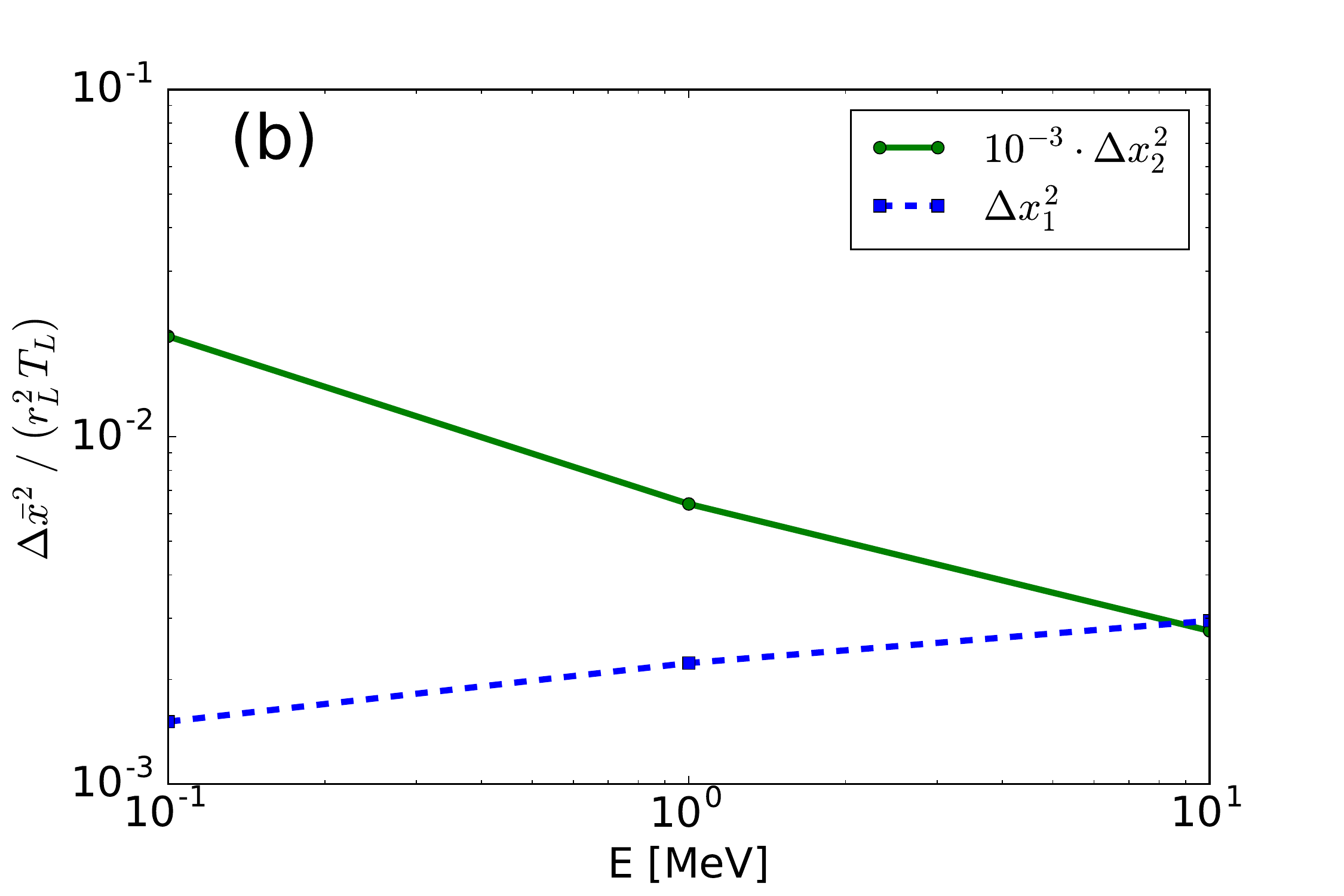}

  \includegraphics[width=.98\columnwidth]{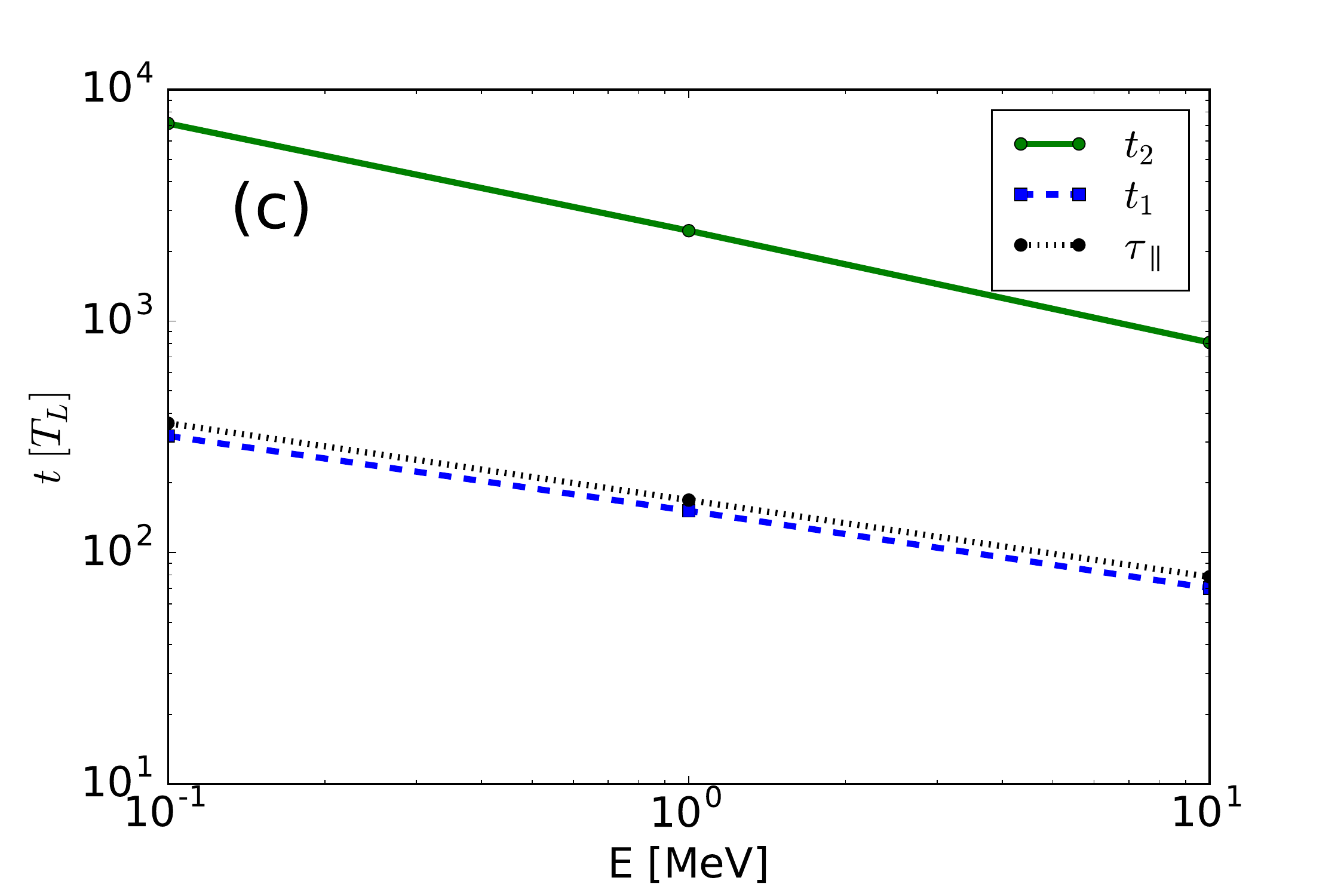}
  \includegraphics[width=.98\columnwidth]{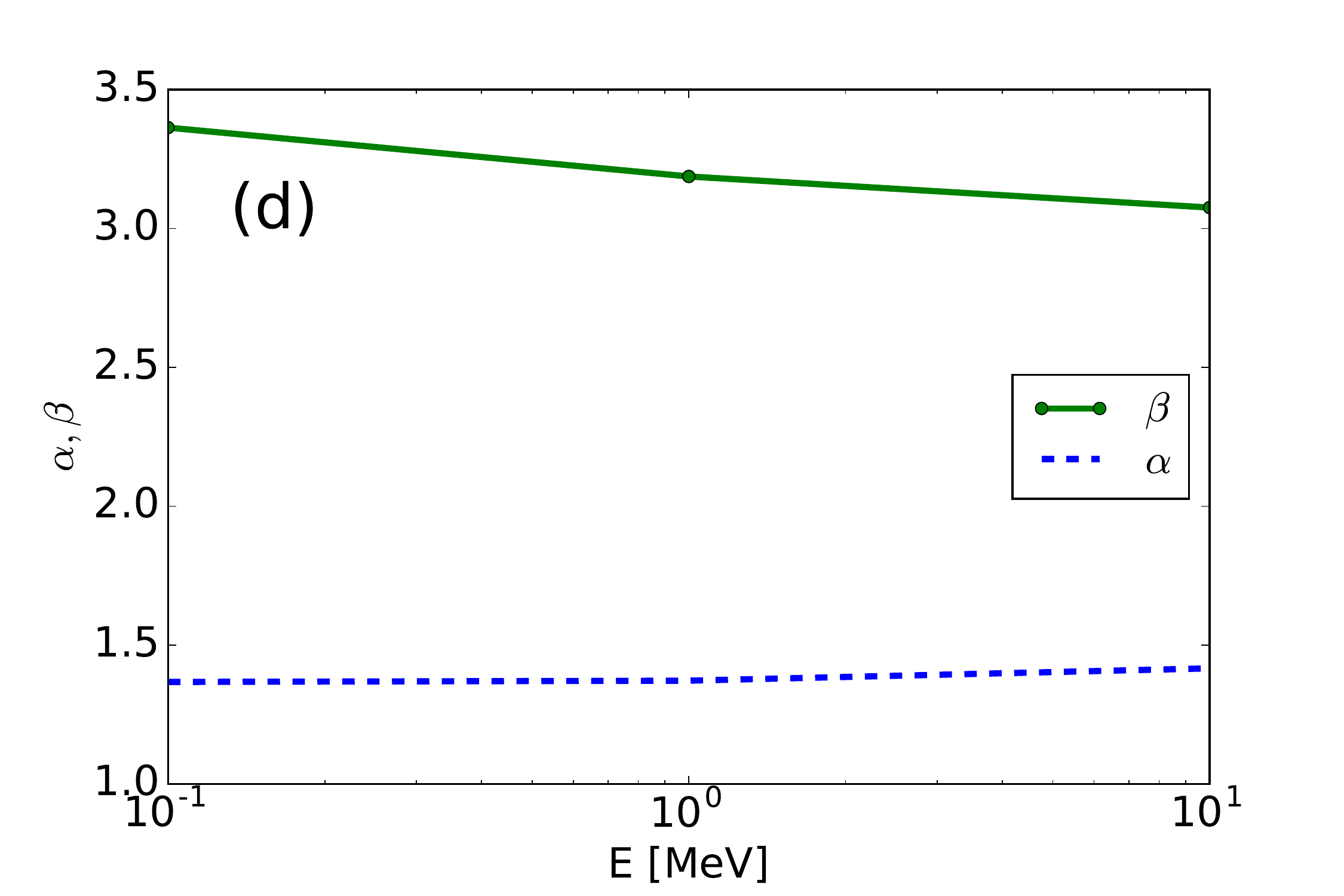}

  \caption{Median displacement versus time for three different proton
    energies in turbulence with $\dbtbt=0.316$, and the corresponding
    fit fo Equation~(\ref{eq:powerfit}). The panels are as described
    in Figure~\ref{fig:t_db2params}.\label{fig:t_Eparams}}
\end{figure*}

We use Eq.~(\ref{eq:powerfit}) to fit the median values of $\dxt$ as a
function of time for different values of turbulence amplitude and
particle energy, excluding times $t<5\, T_L$ from the fitting to avoid
any potential initial non-gyrotropic effects. We choose median instead
of mean as our statistics, as we are interested in the mechanism
causing the transition, and the time when it begins. A mean value is
skewed towards large values due to individual particles having
large displacements. thus, a mean displacement would represent the
extent of the displacement rather than the behaviour of the bulk of
the particles.

In Figure~\ref{fig:t_db2params}, we show the results of our analysis
of the displacement of 10~MeV protons from their initial field lines
in turbulent magnetic field with $\dbtbt=$ 0.0316, 0.0562, 0.1, 0.316
and~0.562, which represent the range of observed turbulence amplitudes
at 1~AU \citep[e.g.][]{Burlaga1976}. In panel~(a), we show the median
displacements and the corresponding fits using
Eq.~(\ref{eq:powerfit}), as a function of time, in units $r_L^2$,
where $r_L=v/\Omega$ is the particle's Larmor radius. The median
values and fits show a three-regime structure observed in
Figure~\ref{fig:scatterplot_time} throughout the analysed $\dbtbt$
range.

In Figure~\ref{fig:t_db2params}~(b) we show $\dxt_1$ and $\dxt_2$,
which quantify the rate of the displacement of the particles from
their field lines in unit time $T_L$ in the first and second diffusive
ranges, respectively. The displacement rate during the first diffusion phase, $\dxt_1$(blue
dashed curve) is a small fraction of $r_L^2$, thus indicating that the
decoupling of the particle from its field line is a slow process
compared to the particle gyration. The displacement $\dxt_1$
depends strongly on the turbulence amplitude, roughly as
$\left(\dbtbt\right)^{2}$.

The displacement rate during the second diffusion, $\dxt_2$,
multiplied by $10^{-3}$ in Figure~\ref{fig:t_db2params}~(b) (solid
green curve), is 3-4 orders of magnitudes larger than $\dxt_1$, and of
order $r_L^2$, which indicates that at time-scales $\gtrsim t_2$ the
particles are fully separated from their initial field lines. The
displacement $\dxt_2$ is roughly proportional to $\dB/B$,
similar to the dependence of field line diffusion coefficient on the
turbulence amplitude in 2D turbulence
\citep{Matthaeus1995}. The decrease of $\dxt_2$ from the
  $\dB/B$ trend at large $\dbtbt$ can be caused by more efficient
  parallel scattering, as can be seen in, e.g., the nonlinear guiding
  centre theory \citep[][]{Matthaeus2003}.

Figure~\ref{fig:t_db2params}~(c) shows the onset time of the
transition phase, $t_1$, and the onset time of the second diffusion
phase, $t_2$, with dashed blue and solid green curves, respectively,
as a function of $\dbtbt$. Both onset times show a
$\left(\dbtbt\right)^{-1}$ dependence on turbulence amplitude, with
$t_2\sim 10\,t_1$. 

We also calculate the parallel scattering timescale,
  $\tau_\parallel=\lambda_\parallel/v$ (black dotted curve), where the
  $\lambda_\parallel$ is the scattering mean free path, obtained as
\begin{equation}
  \label{eq:lambdapar}
  \lambda_\parallel=\frac{3v}{8}\int_{-1}^{1} \frac{(1-\mu^2)^2}{D_{\mu\mu}} \mathrm{d}\mu,
\end{equation}
where $D_{\mu\mu}$ is the quasilinear pitch angle diffusion
  coefficient \citep[e.g.][]{Jokipii1966}, calculated assuming that
  only the slab turbulence contributes to the parallel scattering of
  the particles. For the Kolmogorov turbulence spectral shape used in
  this study, the parallel scattering time depends on the turbulence
  amplitude and particle's Larmor radius as $\tau_\parallel\propto
  \left(dB_\parallel^2/B^2\right)^{-1}\, r_L^{-2/3}$. As shown in
  Figure~\ref{fig:t_db2params}~(c), $\tau_\parallel$ values are close
  to the onset times of the transition phase, $t_1$. This
  implies that the decoupling process of particles from their field
lines may be related to pitch angle scattering of the particles.

In Figure~\ref{fig:t_db2params}~(d), we show the power law indices of
the first diffusion and the transition ranges $\alpha$ and $\beta$,
with the dashed blue and solid green, respectively. The first
diffusion range is super-diffusive, with $\alpha\sim 1.5$, showing
approach to the diffusive limit $\alpha=1$ for higher turbulence
amplitudes. The transition phase (green curve) exhibits a very fast,
super-diffusive cross-field expansion of the particle population from
the initial magnetic field lines.

The Figure~\ref{fig:t_Eparams} shows the median displacement versus
time for proton energies $E=0.1$, 1~and 10~MeV, with $\dbtbt=0.316$,
in the same format as Figure~\ref{fig:t_db2params}. In panel (b) the
rate of the displacement during the first diffusion (dashed blue
curve) depends only weakly on the particle energy, with
$\dxt_1/r_L^2\propto v^{1/3}$. Likewise, panel (d) shows that the
first diffusion range power law index is nearly independent of
the particle energy.

The second phase displacement rate, $\dxt_2$, (solid green curve
in Figure~\ref{fig:t_Eparams}~(b)) decreases as $\dxt_2/r_L^2 \propto
1/v$, or $\dxt_2\propto v$. At time-asymptotic limit
(Equation~(\ref{eq:dx1})), the displacement thus behaves as
$\dxt(t)\propto v t=s$, where $s$ is the distance a particle with
velocity $v$ propagates in time $t$. Thus, the displacement of the
particles during the second diffusion phase is a function of
propagated distance, $s$, only. This indicates that the particle
cross-field propagation during the second diffusion phase is dominated
by the structure of the turbulent magnetic fields rather than the
properties of the particles.

The panel (c) of Figure~\ref{fig:t_Eparams} shows the onset times
$t_1$ and $t_2$ as a function of the particle energy, along
  with $\tau_\parallel$. The onset times scale with energy as
$E^{-1/3}$, with $t_2\sim 10\,t_1$.  As can be seen, the first
  onset time $t_1$ is again very similar ot the parallel scattering
  time, $\tau_\parallel$.

\section{Discussion}\label{sec:discussion}

\begin{figure*}
  \centering
  \includegraphics[width=0.5\textwidth]{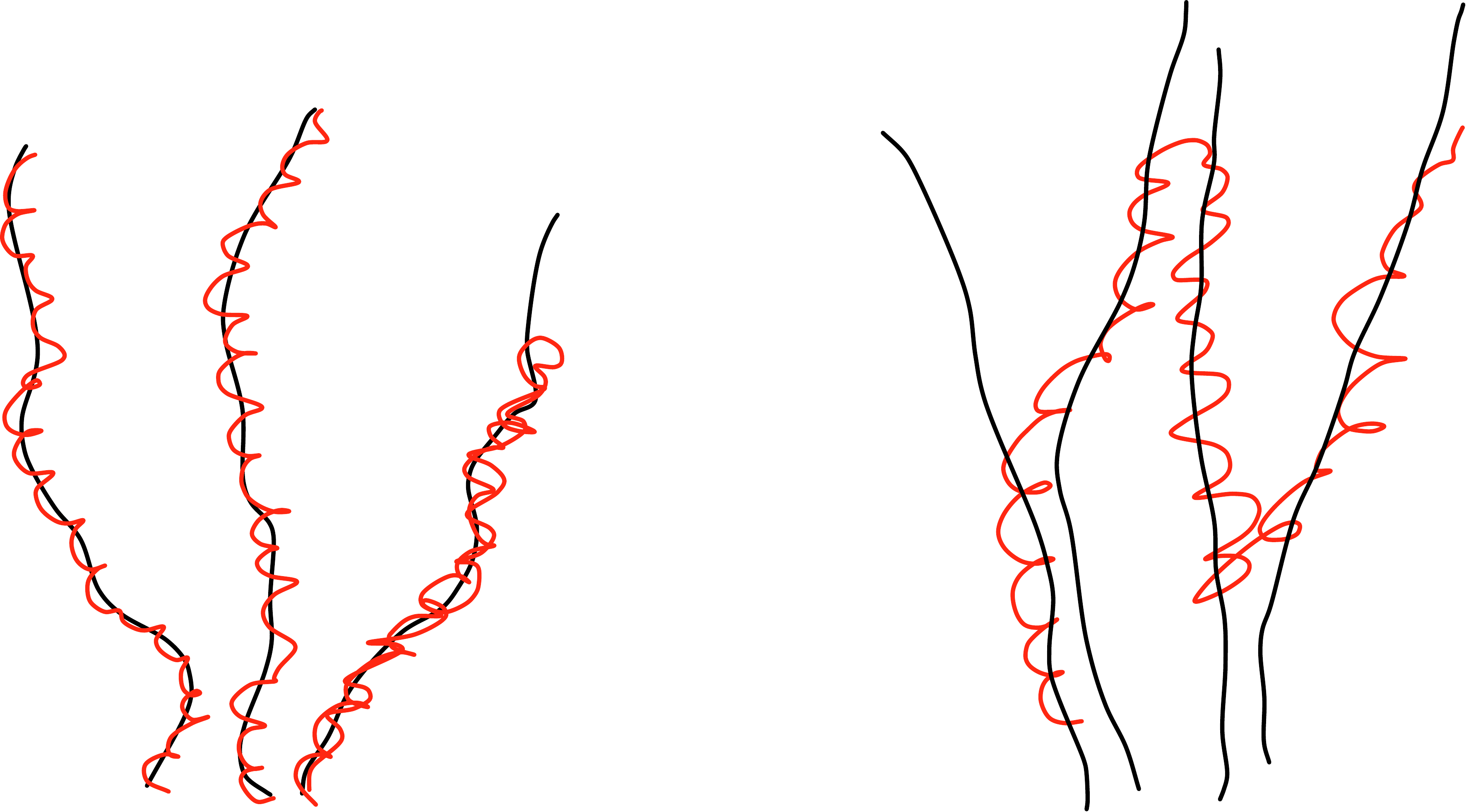}
  \caption{A schematic view of forming of the early (left) and late
    (right) diffusion phases, with the black curves depicting field
    lines and the red curves particle orbits. In the early phase (left
    panel),
    particles remain close to their original field lines, and spread
    across the mean field direction due to the random walk of the
    field lines. In the late phase (right panel), a particle decouples
    from a field line to follow another field line, and as a result,
    propagates across both the mean field and the individual
    meandering field lines.\label{fig:diffusionschema}}
\end{figure*}

Our results show that the propagation of charged particles across the
turbulently meandering field lines can be divided into three phases:
The first diffusion, transition and second diffusion phases. During
the first diffusion phase, the particle displacement from the
meandering field line grows superdiffusively, with the displacement
rate increasing as a function of the turbulence amplitude and particle
energy. As seen in Figures~\ref{fig:t_db2params}~(a)
and~\ref{fig:t_Eparams}~(a), however, during the first diffusion the
displacement of the particles does not exceed the particle's Larmor
radius scale, $r_L$. 
Thus, the particle can be considered as following a single field line
during the first diffusion phase.

It is important to note that this does not imply inhibited particle
propagation across the mean magnetic field during the first diffusion
phase. Rather, the propagation of a particle across the mean magnetic
field is determined by its propagation along a single meandering field
line. The random walk of the field line can cause rapid, non-diffusive
propagation of particles across the mean field direction
\citep{LaEa2013b}. Thus, the particle propagation during the first
diffusion phase follows a scenario depicted in the left panel of
Figure~\ref{fig:diffusionschema}. Recently,
\citet{LaEa2016parkermeand} showed that such a fast cross-field
transport of particles offers and explanation to fast and wide SEP
events with realistic interplanetary conditions already with narrow
source regions.

The first diffusion phase continues until the transition phase onset,
$t_1$, which is of the order of tens to hundreds of gyroperiods for
the particle and turbulence parameters used in this study. For a
10~MeV proton in $\dbtbt=0.1$ turbulence, this corresponds to 0.75
hours, a time in which a particle beam would propagate a distance of
0.7~AU. This implies that the first diffusion range is very
significant for the early propagation of SEPs in the
heliosphere. It should be noted that $t_1$ is much larger than
  the cross-field velocity correlation time obtained from particle
  simulations by \citet{Fraschetti2012perptimesims}. However, their
  method yields the decorrelation time of a particle from an unperturbed
  orbit in a uniform magnetic field, whereas our method yields the decoupling
  time-scale of the particle from a meandering field line.

As shown in Figures~\ref{fig:t_db2params}~(c)
  and~\ref{fig:t_Eparams}~(c), the onset of the transition phase,
  $t_1$, is close to the parallel scattering timescale of the
  particles, $\tau_\parallel$, for the analysed 2D-dominated
  turbulence cases. This could be interpreted as evidence for a strong
  link between the parallel scattering of the particles and the
  particle decoupling from the meandering field lines. However, the
  interpretation cannot be made quite so straightforwardly. In our
  simulations, the particles, all initiated with $v_z>0$, have all
  necessarily experienced pitch angle scattering to have $v_z<0$,
  required for them to return back to the plane $z=z_0$. Thus, during
  the first diffusion, all of the simulated particles have experienced
  pitch angle scattering from the positive to the negative pitch angle
  cosine hemisphere even if their propagation time is much smaller
  than $\tau_\parallel$. Therefore, while
  Figures~\ref{fig:t_db2params}~(c) and~\ref{fig:t_Eparams}~(c)
  indicate that the onset time of the transition phase, $t_1$, is
  close to the parallel scattering time, $\tau_\parallel$, the
  connection between the parallel scattering and the particle
  decoupling from their field lines is likely more complicated than an
  effect due to backscattering of the particles.

The transition phase is rapid and strongly superdiffusive, and
continues until the onset of the second diffusion phase, $t_2$. As
shown in Figures~\ref{fig:t_db2params}~(c)
and~\ref{fig:t_Eparams}~(c), the onset time of the second diffusion
scales as $t_2\sim 10\;t_1$, independent of particle energy and
turbulence amplitude. If we consider the time $t_1$ as the time-scale
of the decoupling of the particle from its field line, the constant
ratio $N=t_2/t_1\sim 10$ can be interpreted as the number of
decouplings taking place until the asymptotic diffusive behaviour in
the second diffusion range is reached. In this interpretation, $t_1$
can be considered as the characteristic time-scale, the ``scattering
time'', for the particle diffusion across the mean magnetic field
line. The scenario of subsequent decoupling of a particle from field
lines leading to particle transport across the mean field is depicted
in the right panel of Figure~\ref{fig:diffusionschema}. 

The transition to the second diffusion phase can be related to
  the recovery of diffusion reported by \citet{Qin2002apjl}, who
  studied the cross-field displacement of particles at all $z$ instead
  of the particles that have returned to $z=z_0$ (our method). They
  noted in their simulations that after an initial fast cross-field
  spreading, the running diffusion coefficient decreased, indicating
  subdiffusion, after which it reached a second diffusion phase. The
  fast spreading seen in the \citet{Qin2002apjl} analysis can be
  understood as particles spreading in space along the meandering
  field lines as depicted in the left panel in our
  Figure~\ref{fig:diffusionschema}, and the subdiffusion due to
  particles backscattering along the meandering field lines
  \citep[compound diffusion: see e.g.][and references
  therein]{Kota2000}. In our simulations, this behaviour is depicted
  by particles remaining in the first diffusion phase, which we have
  quantified in this study. The second diffusion in
  \citet{Qin2002apjl} is likely caused by particles decoupling from
  their field line (right panel in Figure~\ref{fig:diffusionschema}),
  which releases the particles from the the original fieldlines to
  trace the diffusive pattern of the turbulently meandering field
  lines.

As discussed in Section~\ref{sec:results}, the dependence of the
second phase displacement rate, $\dxt_2$, on both the turbulence
amplitude and energy is consistent with the particles diffusing across
the mean field direction in a similar manner as the magnetic field
lines diffuse. Thus, our results are consistent with the recent works
that derive the time-asymptotic cross-field diffusion coefficients
using the statistics of the field line diffusion in the derivation
\citep[e.g.][]{Matthaeus2003,Shalchi2010a, RuffoloEa2012}. However,
the second diffusion is reached only at $t_2$, which is of order
hundreds to thousands of gyroperiods.  For a 10~MeV proton in a
$\dbtbt=0.1$ turbulence, this corresponds to $t_2=$7.5~hours. Thus,
our results suggest that the use of particle transport models where
the cross-field diffusion coefficients are derived at time-asymptotic
limit cannot be justified when modelling the early propagation of SEPs
in the interplanetary space.

It should be noted that the particles being decoupled from their field
lines does not imply that a solution of a diffusive particle transport
equation can be used to describe the particle distribution everywhere
in space. As shown in \citet{LaEa2013b}, the particles at 1~AU from
the injection site spread to a wide cross-field range early in the
event due to field line meandering. While the particles decouple from
the field lines at time-scale $t_2$, the cross-field extent of the
particles at 1~AU is still dominated by the initial spread of the
particles along meandering field lines. As shown in Figure~3 of
\citet{LaEa2013b}, the 10~MeV proton spreading due to decoupling
results in time-asymptotic diffusion behaviour at 1~AU only $\sim$ 20~
hours after their injection in turbulence with $\dbtbt=0.1$
turbulence.

To understand when the particle propagation can be considered as
time-asymptotic, we must understand how the decoupling of the
particles takes place, and how it contributes to the transition to the
time-asymptotic propagation phase. Recently,
\citet{Fraschetti2011perptimetheory} studied the decoupling of
particles from the field lines by deriving a diffusion coefficient for
cross-field propagation of particles due to curvature and gradient
drifts caused by the turbulent magnetic fields. Their first-order
analysis found no contribution from the 2D turbulence to decoupling of
particles from the field-lines, whereas slab turbulence resulted in
subdiffusive decoupling. Thus, their result is not consistent with our
findings, where in the slab+2D turbulence the returning particles
spread from their field-lines superdiffusively, as shown in
Figs.~\ref{fig:t_db2params} and ~\ref{fig:t_Eparams}.

The decoupling of particles from their field lines may also be related
to how the field lines decouple from each other. \citet{Ruffolo2004}
found that neighbouring field lines initially follow each other almost
coherently, with slow diffusive divergence that turns into a
fast spreading at length scale $l_g$, given by
\begin{equation}
  \label{eq:separationcoherence}
  l_g=\frac{\lambda_c}{2}\frac{\dB^2_\parallel}{\dB^2_\perp},
\end{equation}
where $\lambda_c$ is the parallel correlation length of the
turbulence,
\begin{equation}
  \lambda_c=\frac{\pi}{2}\frac{P_{\mathrm{slab}}(k_\parallel=0)}{\delta
    B_\parallel^2}\approx 0.79 L_c.
\end{equation}
In the simulations presented in Figures~\ref{fig:t_db2params}
   and~\ref{fig:t_Eparams}, however,
${\dB^2_\parallel}/{\dB^2_\perp}=1/4$, thus the neighbouring field
lines would decorrelate already at a fraction of parallel correlation
length, thus much shorter than $v\,t_1$
given by our simulations.

\begin{figure}
  \centering
  \includegraphics[width=\columnwidth,trim={0 0 20mm 0}, clip]{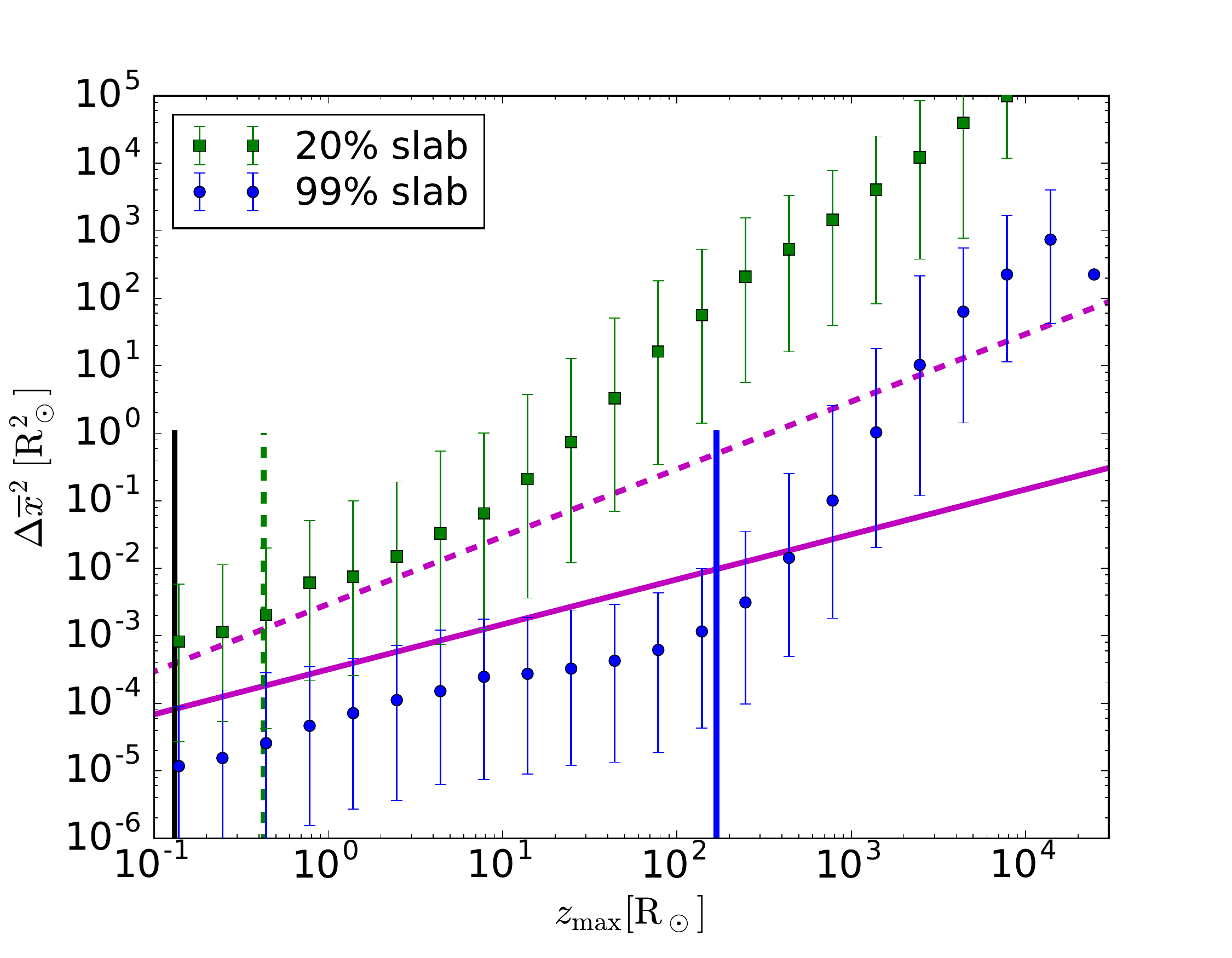}
  \caption{The mean displacement as a function of maximum distance
    along mean field direction for different slab turbulence energy
    fractions, with the error bars depicting the upper and lower
    deciles. For both cases, $\delta B_\parallel^2/B^2=0.112$ and
    $E=10$~MeV. The vertical black line shows the gyroradius of the
    particle. The solid blue and dashed green vertical lines give the
    \citet{Ruffolo2004} field line divergence scale $l_g$ for the
    slab- and 2D dominated cases, respectively. The solid and dashed
    magenta lines show the \citet{Fraschetti2011perptimetheory} result
    for slab and isotropic turbulence,
    respectively.\label{fig:slabish}}
\end{figure}

To further study whether the \citet{Fraschetti2011perptimetheory} and
\cite{Ruffolo2004} formulations can be applied to our results, we ran
additional simulations with slab-dominated turbulence. It
  should be noted that cross-field propagation of charged particles is
  strongly inhibited in pure slab turbulence
  \citep{Jokipii1993,Jones1998}. In addition, the field-line
  separation scale, $l_g$, as defined by
  Equation~(\ref{eq:separationcoherence}), would be infinite in pure
  slab turbulence, indicating absence of strong field-line
  separation. Thus, as we are interested in understanding
  non-negligible cross-field propagation of particles in turbulent
  magnetic fields, we use slab-dominated turbulence with a small 2D
  component instead of pure slab turbulence. Such a slab-dominated
  turbulence mix allows for finite $l_g$, and particle
  propagation is not as constrained as in pure slab turbulence.

In Figure~\ref{fig:slabish}, we show results of simulations of 10~MeV
protons in turbulence with $\delta B_\parallel^2/B^2=0.112$ for 20\%
(green squares) and 99\% (blue circles) slab contributions, with the
error bars representing the lower and upper deciles, respectively. We
present the displacement $\dxt$ as a function of a length scale
instead of time, to gain understanding of the transition process in
terms of~$l_g$. As length scale, we consider the maximum distance the
particle has propagated along the mean field direction,
$z_{\mathrm{max},i}=\mathrm{max}\left\{\left|z_i-z_{0i}\right|\right\}$,
before returning to the plane it was injected at. In our analysis, we
have used mean instead of median square displacement, to obtain better
correspondence with the displacement values predicted by the
\citet{Fraschetti2011perptimetheory} theory. It should be noted,
though, that as we simulate the particles only until their first time
of return to the $z=z_0$ plane, full correspondence with the absolute
values cannot be expected.

In addition to the mean displacement of the returning particles, we
show in Figure~\ref{fig:slabish} the field line divergence scale $l_g$
(Equation~(\ref{eq:separationcoherence})) with the vertical solid blue
line and dashed green line for the slab- and 2D-dominated cases
respectively, and the particle Larmor radius scale with the vertical
black line. The $\Delta x^2$ due to stochastic drifts for slab
turbulence, as given by \citet{Fraschetti2011perptimetheory}, is shown
with the solid magenta curve. It should be noted that our
  simulations are not sufficiently long for analysing the second
  diffusion range by fitting the Equation~(\ref{eq:powerfit}) in the
  slab-dominated case, due to computational limitations. Thus, we will
  concentrate below on analysis of the first diffusion range and the
  onset of the transition phase, and discuss the transition phase only
  qualitatively.

As can be seen in Figure~\ref{fig:slabish}, the first diffusion range
and the subsequent transition phase can be observed for both the
slab-dominated and 2D-dominated turbulence. For the slab-dominated
turbulence, the sub-diffusive trend of the
\citet{Fraschetti2011perptimetheory} result (blue circles) is well
replicated by our simulations. Thus, in the parameter range relevant
to the \citet{Fraschetti2011perptimetheory} work, at scales
  $\ll l_g$ where the field-line separation due to the 1\%
  2D-component is negligible, we find agreement with the previous
theoretical work and our results. The super-diffusive displacement
increase in the 2D-dominated turbulence (green squares in
Figure~\ref{fig:slabish}) seen in our simulations, however, deviates
strongly from the \citet{Fraschetti2011perptimetheory} result. We
suggest two possible explanations for this. One is related to the
field-line divergence scale $l_g$. As shown by the dashed green
vertical line in Fig.~\ref{fig:slabish}, $l_g\sim 3 r_L$ for the
2D-dominated case, well below the start of the transition from the
first diffusion range, at $z_\mathrm{max}\sim 10
\mathrm{R}_\odot$. Thus, the field line de-coherence at short length
scales may influence the particle spreading already during the first
diffusion range, possibly turning the subdiffusive spreading predicted
by \citet{Fraschetti2011perptimetheory} (solid magenta curve)
into the super-diffusive behaviour shown in Fig.~\ref{fig:slabish}.

On the other hand, as speculated by
\citet{Fraschetti2011perptimetheory}, the particle decoupling may be
affected by second-order effects by the 2D turbulence component, which
their theory does not account
for. \citet{Fraschetti2011perptimetheory} present their result also
for isotropic turbulence, shown with a dashed magenta line in
Fig.~\ref{fig:slabish}. As can be seen, it matches the simulation
trend and level of the 2D-dominated case considerably better than
their slab result.

\begin{figure*}
  \centering
  \includegraphics[width=.98\columnwidth]{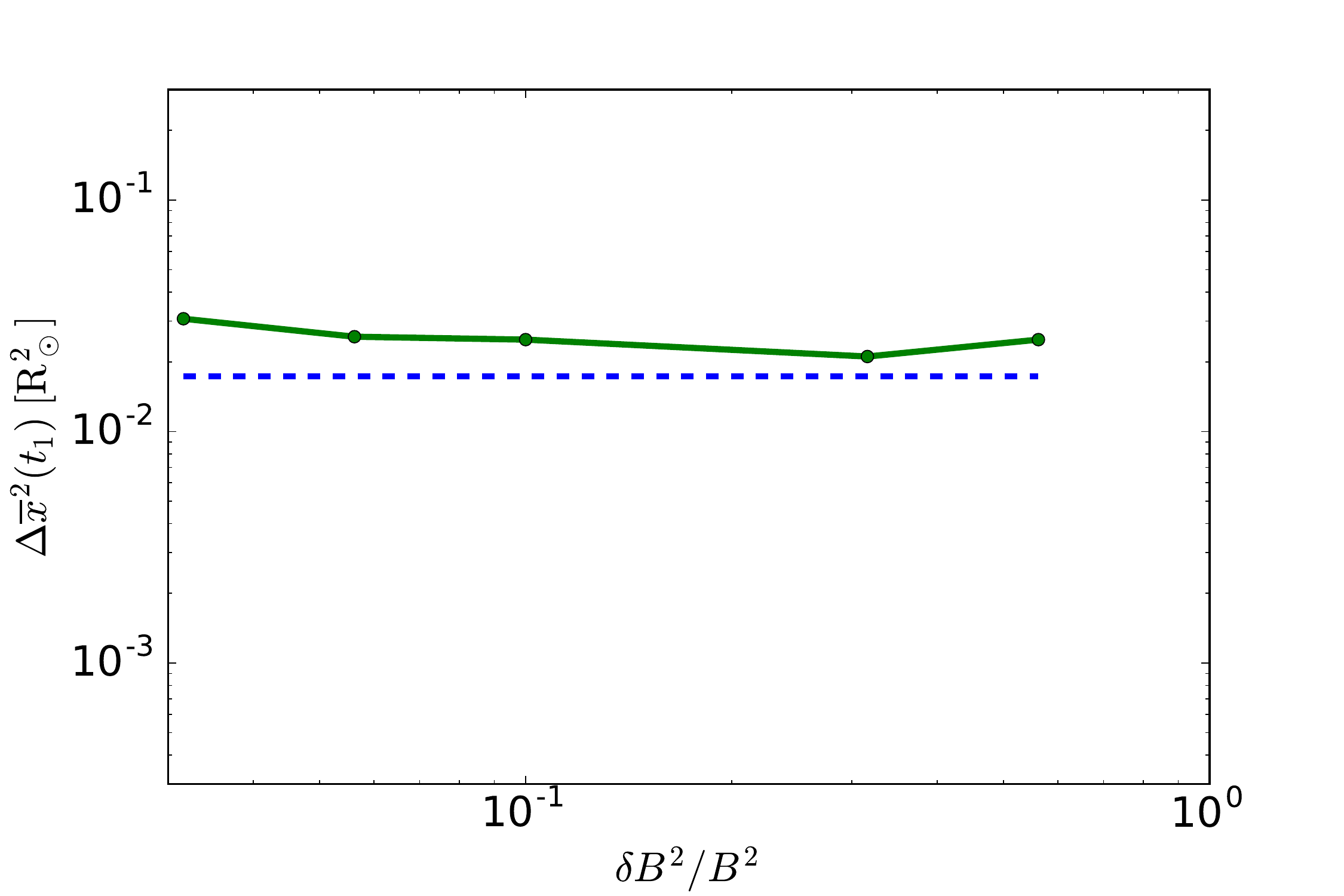}
  \includegraphics[width=.98\columnwidth]{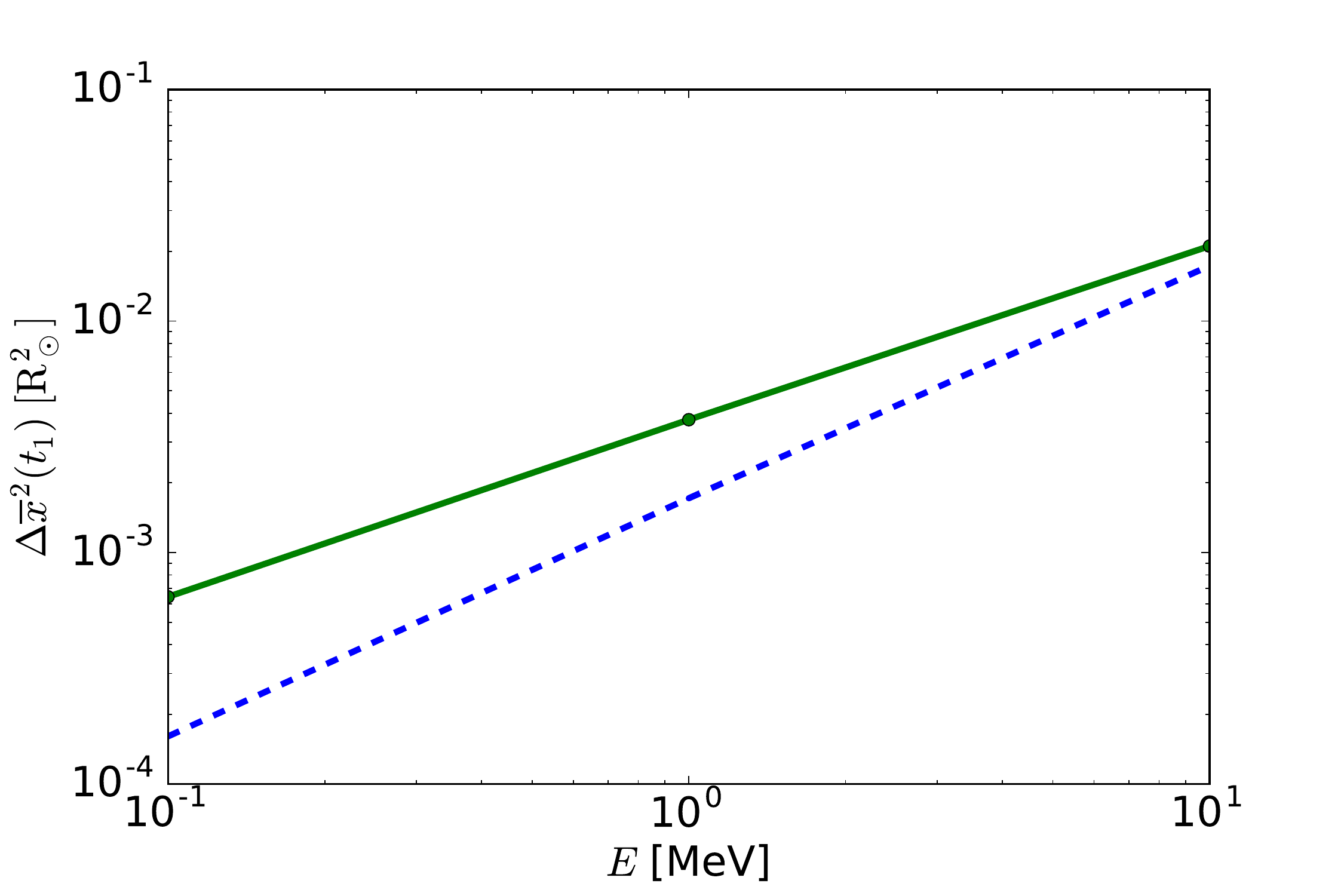}
  \caption{\label{fig:dx2_at_t1} Median displacement of the
    particles at time $t_{1}$, $\dxt(t_1)$ (solid green curve). The
    dashed blue line shows the square of the particle gyroradius.}
\end{figure*}

As shown in Figure~\ref{fig:slabish}, the transition from the first
diffusion to the transition range in the slab-dominated case (blue
circles) takes place for particles that have reached the distance
$z_\mathrm{max}\sim l_g=168\, \mathrm{R}_\odot$. Thus, in the
slab-dominated turbulence, the strong separation of the field lines at
scales $l_g$ \citep{Ruffolo2004} appears to be connected to the
particle decoupling from their field lines. Similar conclusion cannot
be drawn in the 2D-dominated case (the green squares in
Figure~\ref{fig:slabish}): as suggested by the dashed green vertical
line, the field lines are strongly separated much before the onset of
the transition phase, at around $z_{\mathrm{max}}\sim 10\,
\mathrm{R}_\odot$.

Figures~\ref{fig:t_db2params}~(a) and~\ref{fig:t_Eparams}~(a) suggest
a different explanation for the transition of the particle
displacements from the first diffusion range in the 2D-dominated
turbulence. The displacement can be seen to reach the magnitude of the
particle Larmor radius, $r_L$, at the transition onset time, $t_1$ in
all of our simulations with 2D-dominated turbulence. We quantify this
in Figure \ref{fig:dx2_at_t1}, where we compare the displacement of
the particle at the transition onset time, i.e., $\dxt(t_1)$, (solid
green curve), for different simulations, with the particle Larmor
radius $r_L$ (dashed blue curve). As can be seen, at the onset of the
transition range, the particles have moved away by an order of Larmor
radius from their original gyrocenter. It should be noted that in the
slab-dominated case (Figure~\ref{fig:slabish}), the mean displacement
is considerably smaller than the Larmor radius at the time of the
transition, at around $z_\mathrm{max}\sim 200 \mathrm{R}_\odot$. Thus,
the decoupling of the particles from their field lines, as
  defined by the change from the first diffusion phase to the
  superdiffusive transition phase at $t_1$, may be fundamentally
different in slab- and 2D-dominated turbulence.

The role of the particle's Larmor radius as a determining factor for
particle cross-field propagation has been discussed in the context of
electron heat transport in tokamak plasmas by \citet{Rechester1978},
who considered the electrons to be displaced from their field lines by
Coulomb collisions. \citet{RuffoloEa2012} used a similar idea to model
the time-asymptotic diffusion of particles in turbulent plasmas as
ballistic propagation of particles along meandering field lines,
punctuated by decouplings on the time-scale of the parallel scattering
time of the particles. This concept is supported by the close
  similarity of the transition phase onset time $t_1$ and the parallel
  scattering timescale $\tau_\parallel$ seen in the 2D-dominated
  turbulence cases of our study. However, as discussed above, the
  particles analysed in this study have all backscattered in
  field-parallel direction to return to the initial $z=z_0$
  plane. Thus, during the first diffusion, the parallel scattering
  does not efficiently decouple particles from their initial field
  lines.

In addition, the similarity of $t_1$ and $\tau_\parallel$ is
  valid only for the 2D-dominated cases analysed in this study. For
  the slab-dominated scenario we determined $t_1\sim 1000 T_L$,
  considerably larger than the parallel diffusion time,
  $\tau_\parallel=44 T_L$ for $\delta B_\parallel^2/B^2=0.112$ and the
  proton energy of 10~MeV. Therefore, our results indicate that
  scattering alone does not imply a significant displacement of
  particles from their field lines.

Pitch angle scattering may, however, be significant for cross-field
propagation of particles in turbulent magnetic fields. As discussed in
Section~\ref{sec:results}, the transition onset time $t_1$ depends on
the turbulence amplitude and energy in a similar way as the pitch
angle diffusion time-scale $\tau_{\parallel}$, suggesting that faster
pitch angle isotropisation leads to faster transition onset
$t_1$. This may connected to the proportionality of the cross-field
diffusion due to stochastic drifts on $(1-\mu^2)^2$
\citep{Fraschetti2016}, where $\mu$ is the pitch angle cosine. This
dependence indicates that an isotropic distribution would decouple
from the meandering field lines faster from the meandering fields than
an anisotropic one.

It should be noted that in the view of the schematic picture presented
in the right panel of Figure~\ref{fig:diffusionschema}, the dependence
of the cross-field particle diffusion on $\mu$ is not obvious. Strong
cross-field propagation due to field line meandering requires
efficient decoupling of particles from their field lines, and in the
light of the \citet{Fraschetti2016} result, pitch angles $\sim
\pi/2$. On the other hand, the spreading of particles across the mean
magnetic field direction due to propagation along meandering field
lines requires large particle velocities along the field lines,
i.e. $\left|\mu\right|\sim 1$. Thus, pitch angle dependence of the
cross-field particle diffusion particle transport may be more
complicated than the recently discussed proportionality to
$\left|\mu\right|$ or $\left(1-\mu^2\right)$ \citep[see, e.g.,][and
discussion therein]{Droge2010,QinShalchi2014,Strauss2015}. Overall,
our simulations show the importance of understanding the microphysics
of the particle decoupling from their original field lines for
understanding the propagation of particles across the mean magnetic
field in turbulent plasmas.

\section{Conclusions}\label{sec:conclusions}

In this work, we have studied how charged particles spread across the
mean field direction in turbulent magnetic fields superimposed on a
uniform field, by analysing the displacement of a particle from its
initial, meandering, field line instead of the mean field. Our results
show that
\begin{itemize}
\item the particles initially follow their initial meandering field
  lines, over time scales $t_1\sim 50-700\; T_L$, or $10-150$~minutes, for
  a 10~MeV protons in turbulent magnetic fields corresponding to the
  solar wind at 1~AU from the Sun, with $\dbtbt=0.05-0.5$
  \citep{Burlaga1976,Bavassano1982JGR}.
\item the time-asymptotic diffusion, consistent with cross-field
  diffusion dominated by random walk of field lines is reached in
  time-scales $t_2\sim 10\, t_1$, or hours to a day for a 10~MeV
  proton.
\item the transition from the first to the second diffusion range may
  depend on stochastic gradient and curvature drifts, field line
  decoherence and pitch angle scattering of the particles, depending
  on the turbulence parameters.
\end{itemize}

Our results suggest that the first diffusion range, where the
particles stay on their field lines, is significant for SEP event
evolution: 10~MeV protons simulated propagate a distance of 1~AU in
$\sim 60$~minutes, which is of the order of the time-scale a
  particle remains completely bound to its field, $t_1$, in turbulence
conditions corresponding to those of the solar wind at 1~AU. Further,
full relaxation to the time-asymptotic diffusive particle propagation
would be reached in a time-scale of hours to a day. It should be noted
that the turbulence parameters vary radially
\citep[e.g.][]{Bavassano1982JGR}, and also as a function of time
\citep[e.g.][]{Burlaga1976}. The temporal variation will cause also
longitudinal variation of the turbulence parameters, due to the solar
rotation. Also the large-scale magnetic field structure, and the
associated large-scale particle drifts \citep[e.g.][]{Marsh2013}, may
influence the particle decoupling from their field lines. Thus, a full
study including utilising solar wind and turbulence observations and
models is required to understand the SEP propagation in the
interplanetary space in different solar wind turbulence conditions.

Our results indicate that the field line meandering controls the
particle propagation both in the early phases and at time-asymptotic
phases of particle cross-field spreading. However, as depicted in the
schematic view in Fig.~\ref{fig:diffusionschema}, the effect of the
field line meandering manifests itself completely differently at these
phases.  The early cross-field propagation is characterised by
particles following their initial field lines, and is thus
deterministic propagation along stochastic paths. The time-asymptotic
propagation, on the other hand, is characterised by particles
decoupling from their original field lines at time-scales $t_2$, which
causes the particles to random-walk from one random-walking field line
to another. 

The mechanism of the early time particle cross-field propagation
presented in our study provides also a possible explanation for the
SEP intensity dropouts. These dropouts, observed in some SEP events
\citep[e.g.][]{Mazur2000}, imply strong cross-field gradients in
spatial SEP distribution in these events. While such gradients would
be smoothed by the time-asymptotic cross-field diffusion
\citep{Droge2010,Wang2014}, our simulations show that the cross-field
propagation of the particles across the meandering field lines is
negligible during the first diffusion phase, enabling the intensity
dropouts to persist.

The mechanism behind the transition phase between the first and second
diffusion ranges remains unclear, and requires further
study. Our results show that the first diffusion phase and the
  transition phase of the particle cross-field transport exists both
  for 2D and slab-dominated turbulence, which suggests that the
  transition is a general feature in early cross-field propagation of
  particles in turbulent magnetic fields. We have identified
potential mechanisms through comparison with the stochastic drift
diffusion theory \citep{Fraschetti2011perptimetheory}, the field line
decoherence \citep{Ruffolo2004}, parallel scattering timescale
  of the particles, and the distance the particle deviates from its
field line before the transition commences. However, the
  relative contribution of different processes appears to depend on
  the composition of the turbulence. In a future work, we will study
the interplay between different phenomena contributing to particle
cross-field propagation by using guiding centre simulations that
include the relevant physics and comparing them to full-orbit
simulations as well as theoretical results. Such a study would be
capable of improving our understanding on not only the transition
stage of the charged particle cross-field propagation, but also how
the time-asymptotic diffusive behaviour is formed.

\acknowledgements{ TL and SD acknowledge support from the UK Science
  and Technology Facilities Council (STFC) (grants ST/J001341/1 and
  ST/M00760X/1), and the International Space Science Institute as part
  of international team 297. Access to the University of Central
  Lancashire's High Performance Computing Facility is gratefully
  acknowledged.}

%\bibliographystyle{apj}
%\bibliography{ms}

\end{document}